\documentclass[twocolumn,showpacs,preprintnumbers,amsmath,amssymb]{revtex4}
\usepackage{graphicx}
\usepackage{dcolumn}
\usepackage{bm}

\newcommand{\eq}[1] { Eq.~(\ref{#1})}
\newcommand{\be}{\begin{equation}} 
\newcommand{\ee}{\end{equation}} 
\newcommand{\bea}{\begin{eqnarray}} 
\newcommand{\eea}{\end{eqnarray}}

\begin{document}
\title{Heat and 
  super-diffusive melting fronts in  unsaturated porous media}
\author{Eirik G.\ Flekk{\o}y${}^{1,2}$, Alex Hansen${}^{3}$ and  Erika Eiser ${}^{3}$ }
\email{flekkoy@fys.uio.no}
\email{alex.hansen@ntnu.no}
\email{erika.eiser@ntnu.no}
\affiliation{
${}^1$PoreLab, Department of Physics, University of Oslo, 
NO--0316 Oslo, Norway\\
${}^2$PoreLab, Department of Chemistry, Norwegian University of 
Science and Technology, NO--7491 Trondheim, Norway\\
${}^3$PoreLab, Department of Physics, Norwegian University of 
Science and Technology, NO--7491 Trondheim, Norway }
\date{\today {}}
\begin{abstract}
When water is present in a medium with  pore sizes in a
range around
10nm  the corresponding freezing point depression
will cause long range broadening of a melting front. Describing the
freezing-point depression by the Gibbs-Thomson equation and the
pore size distribution by a power law, we derive a non-linear
diffusion equation for the fraction of melted water. This equation
yields super-diffusive spreading of the melting front with a diffusion
exponent which is given by the spatial dimension and the exponent
describing the pore size distribution. We derive this solution
analytically from energy conservation  in the limit where all the
energy is consumed by the melting and explore the validity of this
approximation numerically. Finally, we explore a geological application of the
theory to the case of one-dimensional sub-surface melting fronts in
granular or soil systems. These fronts, which are produced by
heating of the surface, spread at a super-diffusive rate and affect the
subsurface to significantly larger depths than would a system without
the effects of freezing point depression. 
\end{abstract}
\maketitle
\section{Introduction}

\begin{figure}[h!]
\begin{center}
\includegraphics[width=0.90\columnwidth]{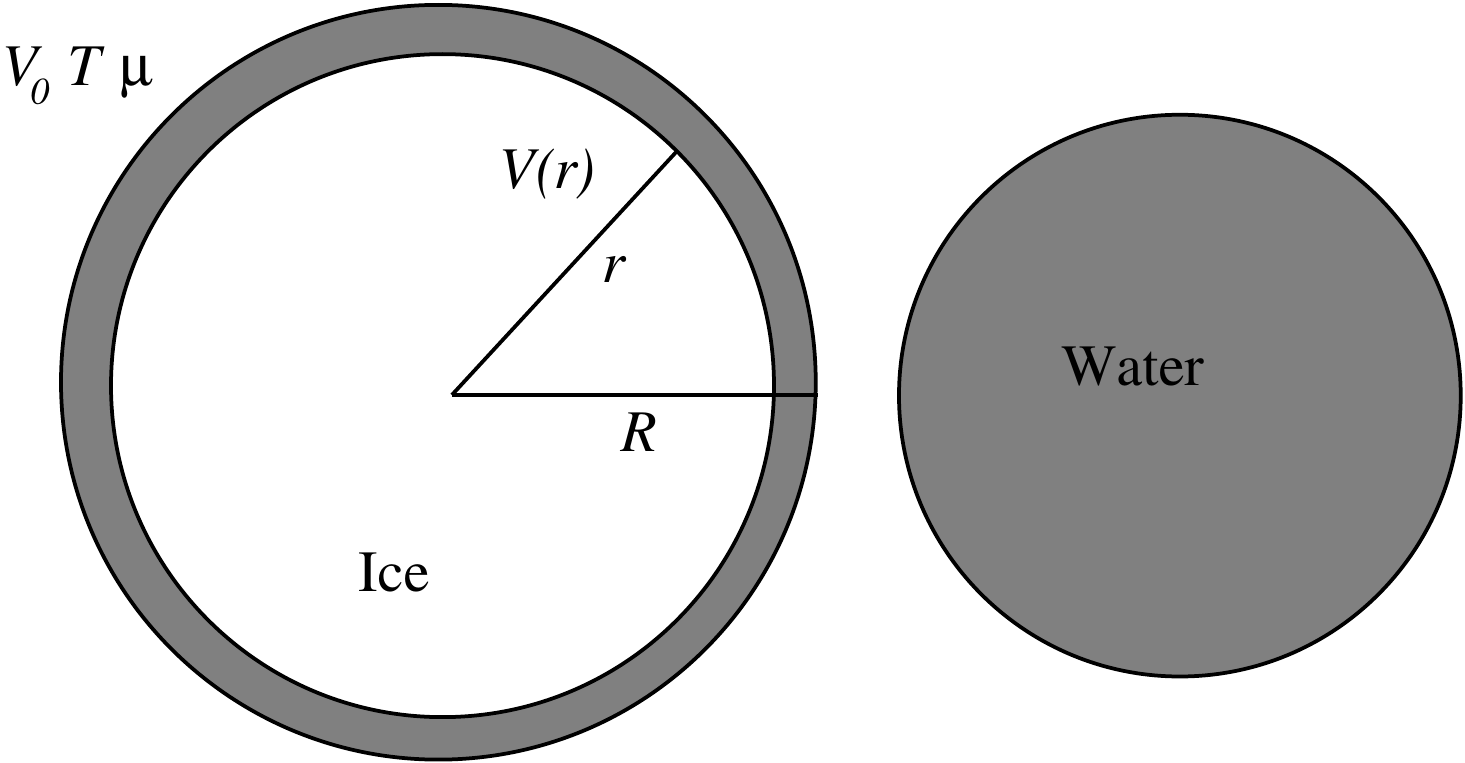}
\caption[]{Left: A pore of total volume $V_0$ containing a liquid film
and a core of ice. Right: A smaller pore containing liquid water at
the same temperature.}
\label{jhgjgayt}
\end{center}
\end{figure}

\begin{figure}[h!]
\begin{center}
\includegraphics[width=0.90\columnwidth]{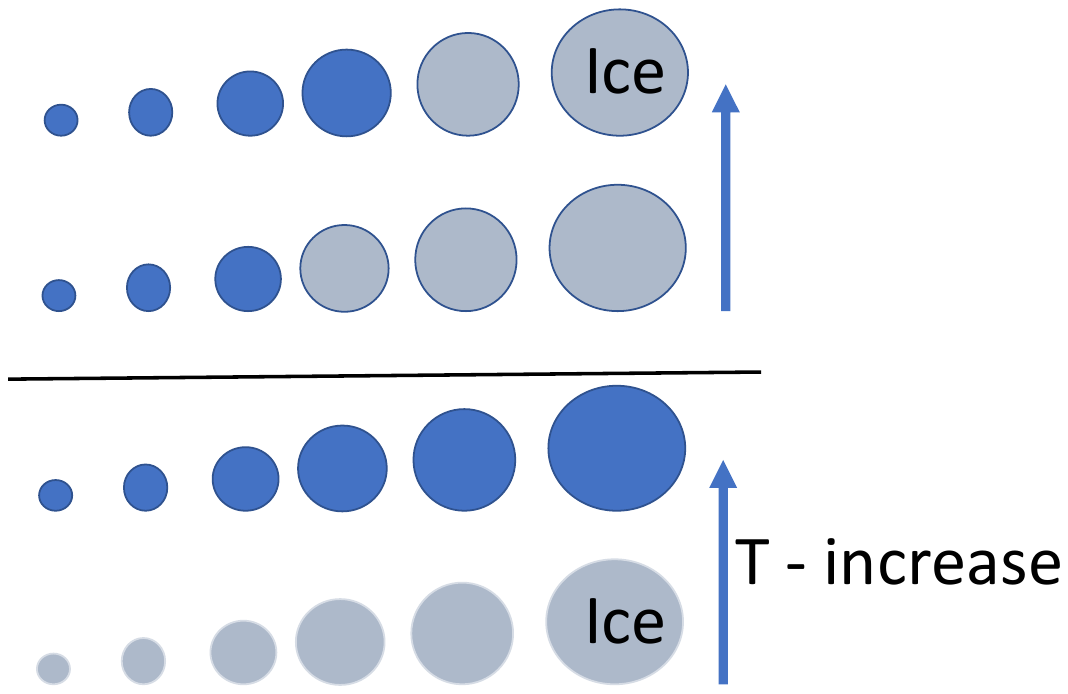}
\caption[]{Pores containing water. Ice melting with (upper figure) and without (lower figure)
  the freezing point depression. Ice is shown in gray, liquid water in
blue.}
\label{jhkuytgjgayt}
\end{center}
\end{figure}

Water residing in $\sim$ 10 nm pores will stay in the liquid state
at temperatures well below the bulk freezing point.
Such freezing point depression is caused by the Gibbs-Thomson
effect, which in a porous medium with a range of pore sizes, will
cause residual amounts of liquid water in small pores while water in
the larger pores freezes.
The frozen state of a single pore is illustrated in   figure  \ref{jhgjgayt}, where a
pre-melted layer of liquid water is assumed to be present, while the
situation where pores of different sizes coexist, is illustrated in 
figure  \ref{jhkuytgjgayt}.
Several experimental  studies of the freezing point depression in small pores have been carried out,
showing that quantitatively, the effect depends 
on such factors as salinity \cite{watanabe02},
 wetting properties \cite{moore12} and the   pore
geometry \cite{marcolli14,campbell18,lazarenko22}.

The equilibrium states of frozen systems has been
studied experimentally in a both natural \cite{watanabe02} and
synthetic media, such as cylindrical silica
nanopores\cite{findenegg08} of controlled sizes in the 2-10 nm range.
However, much less has been learned about the non-equilibrium processes of heat
propagating through such systems where only a fraction  of the ice melts.
When sufficient amounts
of water is present at the right temperature, the energy required for
this melting will dominate the energy balance, i.e. the latent heat is
larger than the energy needed to change the temperature due to the
heat capacity.
With different pore sizes present the heat may be consumed by melting
only in a narrow range of these sizes (see figure  \ref{jhkuytgjgayt}).
This causes an enhanced spreading of the heat as well as the
fraction $C$, of melted water. We will show that this fraction may
spread in a  super-diffusive manner when the pore size distribution is given by
a power law. By comparison, a melting front in  medium where all the pores have the
same size and  melts at the same temperature, will  stop abruptly  at
the point in space where  the available energy is consumed, and
thus have no long tail.
Superdiffusion is characterized by the fact that second spatial moment
of the water fraction $C$,  increases with time
$t$ as  $\sim t^{\tau }$ with the exponent $\tau >1/2$, the 
normal  diffusion value being $\tau=1/2$.
This behavior may arise in physical, biological  or
geological systems;  examples include  Levy flights \cite{bouchaud90,gosh16},  particle 
motion in random potentials or the seemingly random paths of objects
moving in turbulent flows\cite{richardson26,schlesinger87}.

In addition to the shift in the equilibrium freezing point itself,
there may be an effect of metastable states that cause super-heating
or super-cooling.
In order to address this question we
discuss qualitatively how the Gibbs-Thomson effect  may be 
modified by nucleation barriers as well as the pore geometry and shapes of
the ice. However, since melting process is generally less 
affected by nucleation barriers and alternative nucleation pathways \cite{marcolli14,marcolli17,campbell18} 
than the freezing process,  our theory is formulated for melting fronts and
proceeds on the basis that metastable states may be neglected \cite{hu2003free,joost85}. 

We show that when
the porous medium has a power law pore size distribution, the fraction
of liquid water satisfies  a non-linear diffusion equation.
Solving this equation analytically we proceed to demonstrate that
this  results in a super-diffusive-, and, in some cases, even
hyper-ballistic spreading of the heat and liquid
concentration. The diffusion exponent is  given in terms of
the exponent governing the pore size distribution and the dimensionality. 

These results may be of relevance for the modelling of  melting in such
environments as tundras.
We therefore apply the model result to explore potential consequences
for the depths at which Gibb-Thomson effect
may affect the melting of ice in such contexts. Given the above assumptions
the depths at which the ice fraction is perturbed,
may be up to a factor 10 larger than without the effect of freezing point depression.
We also show numerically that this effect survives, even with realistic values for energy consumed by the heat
capacity of the water and the solid medium.

The paper is organized as follows: In the theory section we introduce
the standard thermodynamics of the Gibb-Thomson effect, deriving the
expression for the freezing-point depression.
Following the discussion of the  equilibrium states, we discuss the
assumption of a power law distribution for the pore sizes before we
turn to the consequences for time-dependent equation that governs the
evolution of the melted water fraction and obtain its solutions in different
spatial dimensions.  Finally, we interprete these results in an
assumed geological scenario where a melting front is caused by
surface heating which leads to a long-range, super-diffusive spreading
of the melting front.   


\section{Theory}

In the following we obtain the volume fraction of liquid water as a
function of temperature for a porous medium with a given pore size
distribution and  water/ice saturation $C_0$.   For this purpose we
need the freezing point as a function of pore size.

It is a general fact that most water bearing 
solids, or even  ice itself \cite{elbaum91}, will have a pre-melted liquid  layer 
\cite{wilen95,israelachvili11} of a thickness $\sim$nm, as illustrated
in figure \ref{jhgjgayt}. 
While the thickness of the melted films varies with the interaction  
energy between the water molecules and the walls \cite{moore12}, the  
existence of the film is quite insensitive to the corresponding  
wetting properties of the wall.

Being interested in pores on the nano- to micrometer scale we
will assume that the chemical potential $\mu$ is constant over the
pores. This is justified by the fact that diffusion is fast on these
scales, and so the water will quickly equilibrate to the chemical
potential of the surroundings. This will be assumed to be the  case whether the pore
is open to the surrounding pore volume, or not.
The situation is illustrated in figure  \ref{jhgjgayt}. In this case
a body  of ice  will adjust its  volume $V(r)$ so as to minimize the
Landau-, or grand canonical potential $\Omega$ in
equilibrium.
We will not consider the case where the increase in specific volume of
the water during freezing leads to significant pressure changes.
So, the theory is limited to the cases where there is some freedom for
the water to expand or be absorbed, as is generally the case in unsaturated or
unconsolidated porous media with boundaries that are open to the
surroundings.

\subsection{Freezing point depression and the thermodynamics of the Gibbs-Thomson effect in spherical pores}

The  net energy
effect of introducing a liquid layer between a solid (or vapor)  and ice
may be
described by the Landau free energy
\be
\Omega_{VW} =A_0 A_H/(12\pi d^2) ,
\label{ufuyt}
\ee
where $A_0$
is the surface area, $d$ the liquid layer thickness,  and we have introduce the Hamaker constant $A_H$ $\sim
10^{-20}-10^{-19}$ J (albeit with an unconventional sign so as to keep
$A_H$ positive).

\begin{table}
  \centering 
\begin{tabular}{||lc|r||} 
    \hline 
$   \kappa_{ice}$ & = 2.3  W$/$(mK) & thermal conductivity of ice  \\
    \hline 
$   \kappa_{w}$ & = 0.6 W$/$(mK) &thermal conductivity of water \\
    \hline 
$   \kappa_{b}$ &  $\sim $ 1 W$/$(mK) &typical thermal conductivity of  clays \\
    \hline 
  $ \lambda $ & =  0.33    MJ$/$kg &latent heat of fusion for water \\
    \hline 
  $ \sigma $ & = 0.033 N$/$m &water-ice surface energy per unit area \\
    \hline 
$   \rho_i$ & =  917  kg$/$m$^3$& mass density of H$_2$O ice \\
    \hline 
 $  c_{ice}$ & =  2.3  MJ$/$Km$^3$&  heat capacity of ice  \\
    \hline 
  $ c_{w}$ & =  4.2  MJ$/$Km$^3$ &heat capacity of water \\
    \hline 
  $ A_H$ & =  $10^{-20}$-$10^{-19}$
           J & typical values of Hamaker constant \\
           \hline 
\end{tabular}
 \label{material}
\caption{  Material constants  }
\label{table}
\end{table}

Adding the free energy of the ice-water interface $
\sigma A(r)$, where $A(r)$ is the area of this interface,
  where 
$\sigma$ is the ice-water surface energy per unit area,
to the
energy of the pre-melted layer given in \eq{ufuyt}
yields the
total
free energy
\be  
\Omega = \sigma A + \frac{A_H A_0}{12\pi (R-r)^2} + \Omega_0  , 
\label{kiuyg}
\ee  
where the bulk free energy $\Omega_0$ is independent of
the interface contributions. Since in general $\Omega = - PV$, the
combined potential for both the liquid and ice is
\be
\Omega_0 = -P_iV_i - P_w (V_0 - V_i), 
\ee
where the ice pressure $P_i$ and water pressure $P_w$ will in general
differ.

At the bulk melting temperature $T_m=$ 273 K, there will be no change
in $\Omega_0$ under a change in $V_i$ when $\mu$ and $T$ are kept fixed,
so, using the fact that $\Omega_0 = E - TS - \mu N$,  we can write
\be
0= d\Omega_0 = dE - T_m dS - \mu dN
\ee
where $E$, $S$ and $N$ are the total ice-water energy, entropy and
molecule number respectively. The heat needed to melt a volume $-dV_i$
of ice is
$\rho_i \lambda dV_i$,  where $\rho_i$ is the ice mass density and 
$\lambda$
the latent heat per unit mass. Using this  the above equation may also be written
\be
-\rho_i \lambda dV_i = T_m dS=  dE - \mu dN .
\ee
Since $\rho_i$ and $\lambda$ changes very little over a modest
temperature variation, we may also get $d\Omega_0$ in the case where
$T\neq T_m$ by writing
\begin{align}
  d\Omega_0 &= (dE - \mu dN) -  
              \frac{T}{T_m}T_m
  dS \nonumber \\ &\approx - \left(   1  -\frac{T}{T_m}   \right)\rho_i \lambda dV_i ,
\end{align}
which shows that the free energy change due to an ice volume increase
is negative below the bulk freezing point.
Integrating  from $V_i=0$  where $\Omega_o = -P_w(\mu , T) V_0
$, yields
\be
  \Omega_0 \approx - \left(   1  -\frac{T}{T_m}   \right)\rho_i \lambda V_i  -P_w V_0,
  \label{bulk}
  \ee
which, when inserted in \eq{kiuyg} gives
\be  
 \Omega = \sigma A + \frac{A_H A_0}{12\pi (R-r)^2}   - \left(   1  -\frac{T}{T_m}   \right)\rho_i \lambda V_i  -P_w V_0. 
\label{kiuiyg}
\ee  
The change in this energy as $r$ is increased from $r=0$ is  
\be  
\Delta \Omega = \sigma A  - \left(   1  -\frac{T}{T_m}   \right)\rho_i \lambda V_i  
+  \frac{A_H 
}{12\pi}
\left(
  \frac{A_0}{ (R-r)^2}
  -  \frac{A_0}{ R^2}
\right) .
\label{kiuiyg}
\ee  
 
\begin{figure}[h!]
\begin{center}
\includegraphics[width=0.90\columnwidth]{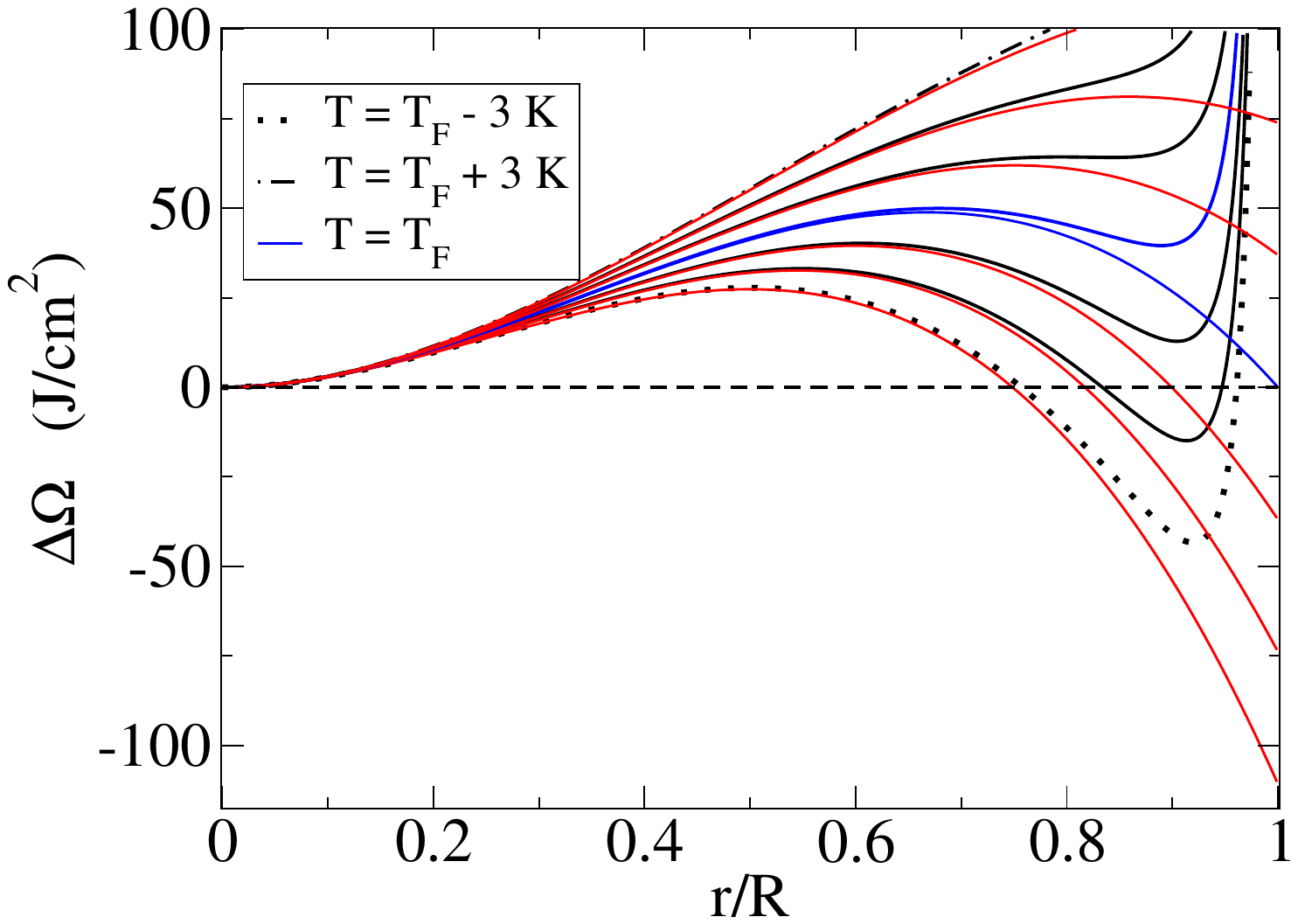}
\caption[]{The Landau free energy per unit area in  a pore of radius  
  $R=10$ nm with a  
  surface layer of water as a function of the radius $r$ of the ice  
  volume at temperatures around $T_F=$ 264 K. The
  black  curves show the free energy with $A_H=10^{-20}$ J, while  the  
red curves show the free energy with $A_H=0$. For the blue curves $T=T_F$. }
\label{jhgayt}
\end{center}
\end{figure}

The equilibrium value of the  ice radius is given by the global minimum of  $\Delta \Omega$,
which, for sufficiently small $R$ values, will be at $r=0$, that is, for
the complete liquid state.  Above this critical pore size, the minimum
will be at $r_m \lesssim R$, a value that is given by the equilibrium
thickness of the surface melted layer. As may be noted from figure  \ref{jhgayt} 
this minimum does not change  
much with the pore size $R$. When $ \Omega (r=0) > \Omega (r_m) $
there will still be ice. 
The condition for complete melting  is
that $ \Omega (r=0)< \Omega (r_m) $, which yields
the freezing point depression.

Taking  $r=R$,  $A_H=0$ gives the free energy change in passing from
a liquid to a fully frozen pore 
\be  
\Delta \Omega =
\sigma A  - \left(   1  -\frac{T}{T_m}   \right)\rho_i \lambda V_i , 
\label{kiuiygi76t}
\ee    
where $A=4\pi R^2$ and $V_i=(4/3)\pi R^3$. 
The condition $\Delta \Omega =0$ implies that a
pore of radius $R$ will freeze at a temperature $T=T_F$ given by 
\be  
T_F(R)=T_m \left( 1 - \frac{3 \sigma}{\rho_i \lambda R}  \right) .  
\label{ytfuapp}
\ee  
This is is the standard expression for the Gibb-Thomson effect.
For a cylindrical pore, the geometrical factor of 3 must be replaced 
by 2. In the following we shall use the value 3.  Note that, due to 
the tendency of the surface tension to minimize the interface area, these smooth 
geometrical  shapes will also be relevant in more complex pore
geometries.

\subsection{Corrections to the Gibbs-Thomson effect due to nucleation 
  barriers}

So far we have ignored the time it takes for a metastable state to be
replaced by the equilibrium state, implicitly assuming that the
system has had time to reach the overall minimum state for the free
energy.  This is in general not the case as some metastable states may
be very long-lived, a phenomenon that is quantified in classical
nucleation theory \cite{hanna06,frenkel2001} which is based on the
probability that a free energy barrier is traversed by the thermal
activation energy $k_BT$. 
Moreover, the stability against melting may be very different from the
stability against the reverse process of freezing. 
It is generally much more difficult to superheat a solid than to 
supercool a liquid \cite{hu2003free,joost85}. 
Superheated crystaline solids have only been observed in some rather singular
cases where the heated region is along a single crystal plane or the
crystals are confined inside a non-melting matrix
\cite{daeges86,grabek92}. 
The existence of supercooled liquids on the other hand only requires
the absence of nucleation sites. 

Assuming our pre-melted surface layer of water, there is no extra energy cost (nucleation barrier) in 
forming new  liquid-ice surface  during a melting process.
Yet, there will be a nucleation barrier that must be crossed 
during melting when the  temperature is $T\approx T_F$. The reason for
this is that when melting happens around $T\approx T_F<T_m$, the bulk
free energy has to increase while the surface energy is decreased, and
so as $r$ decreases from a value around $R$ in 
figure \ref{jhgayt},
the free energy initially increases. As a result
there is a free energy barrier against melting as well as
freezing. This fact implies the possible existence of solid ice that
is superheated relative to its depressed freezing point $T_F$.

Using nucleation theory  it is possible to estimate the lifetime of
these metastable states as $\propto \exp ( (\Delta \Omega (r_{max}) -
\Delta \Omega (R))/(k_BT))$, where the free energy  $\Delta \Omega
(r)$ is shown in figure \ref{jhgayt} and takes its maximum at
$r=r_{max}$.
Requiring that the lifetime be within a realistic range it is possible to show that the
melting temperature must be raised above $T_F$ by an amount that
corresponds to a reduction of the freezing point depression $T_m-T_F$ by $\sim$
20 \% for pore sizes greater than $\sim $ 1nm. 

Furthermore,  it may be shown  that nucleation barriers are
significantly more  influential  during freezing (supercooled
liquid).  In this case, however, nucleation pathways other than ice
forming as a spherical crystal are likely to dominate, as has been
shown for the case where ice nucleates in pockets or corner geometries \cite{marcolli14,marcolli17,campbell18}.

In the following we will consider melting
on the basis that metastable states may be neglected, in spite of the
fact that   there is a nucleation 
barrier to be passed both for the melting and freezing transition in 
isolated pores. For melting this assumption implies that there may be
quantitative corrections to the  depression $T_m-T_F$ by $\sim$ 20
\%, 
which are ignored.

\subsection{ Heat in a nano-porous medium with partially frozen water}

Having dealt with the equilibrium problem of the freezing point
depression we now turn to investigate the non-equilibrium effects of
this phenomenon in the context of a nano-porous material. We shall
consider a melting front, for which the shift in melting temperature
is small,  and so the shift in the freezing point
depression  will not be applied.
Note, however, that a freezing front may differ significantly from the
melting front through the possible existence of metastable pockets of
supercooled liquid. 

The pore size distributions may be estimated through nitrogen absorption \cite{sing01},
electron microscopy, or mercury injection experiments and measurements
of the heat capacity variations with temperature when there is water present \cite{tombari06}. For silts, clays and synthetic
media made of glass powders \cite{watanabe02}
they may yield distributions that extend down at least to the
nm-scale.
Freezing and melting of water confined in silica nanopores has been 
observed down to pore sizes of 3 nm\cite {findenegg08}. 

The distributions may be given in terms of a relative volume fraction
per unit length $g(R)$ so that $\int dR g(R)= \phi$, the porosity of
the medium.
Our main assumption is that this distribution may be approximated with
a power-law above a minimum cut-off length $R_{min}$,
\be
g(R) = N(R-R_{min})^{\beta}
\label{distrjih}
\ee
where $N= (\beta +1 )\phi /(R_{max}-R_{min})^{\beta +1}
$ is just the normalization.

Mercury
intrusion experiments are  challenged by the fact that
high injection pressures may crush or deform the smallest pores. Yet,
in rigid materials, such as cement, the technique may be used to
measure pores down to $R_{min} \sim$ 1 nm \cite{zhu19}.
In order to cover the smaller pore ranges nitrogen adsorption
techniques are often better \cite{sing01}. 
Zhao et al. \cite{zhao17}  measured pore size distribution for porous sandstone from the 
Ordos basin  by mercury injection, finding $g(R)$-distributions that
are well described by  $R_{min} \sim$ 10 nm and $\beta=1 $ over one to
two decades in pore sizes. 
Using  N$_2$ adsorption techniques on   porous glass powders, 
  Fujinomori soil  and Bentonite clay  
Watanabe et al. \cite{watanabe02} found
$g(R)$-distributions  where $R_{min} \approx$ 1 nm, $R_{max}\approx$ 3-4 
nm and $\beta=$ 1-2.
  Park et al. \cite{park18} measured pore size distribution of some
natural sediments
 using mercury intrusion porosimetry. Different sediments
produced $R_{min}$-values from 1-100nm. with distributions that could
be described by a $\beta \approx 2$ power law over roughly a decade. There is
thus a range of natural and synthetic materials that seem to fulfill
the assumed pore size power law distribution over an adequate range
of length scales.

In a medium that is described by \eq{distrjih} all the pores are frozen when
\be 
T=T_{min} =T_m - T_m \frac{r_0}{R_{min}},
\label{jkhbrwa}
\ee
where we have introduced the length $r_0=3 \sigma/(\rho_i \lambda)
\approx 3.3$nm.
Correspondingly, there is an upper temperature
\be 
T_{max} =T_m - T_m \frac{r_0}{R_{max}},
\ee 
where all the pore water is melted. 

The initial filling fraction of water in the pores $C_0$ gives
the total water (ice or liquid) fraction $\phi C_0$.
The fraction of liquid water $C(T)$, is the fraction contained in the
pores that are so small that they have not frozen. These pores have 
sizes less than
\be
r(T) = \frac{3 \sigma}{ \rho_i \lambda (1-T/T_m)}.
\ee
This means that when $T_{min}<T<T_{max}$
\begin{align}  
  C(T) & = C_0 \int_{R_{min}}^{r(T)} dRg (R) \nonumber \\
&= N  C_0 \int_{R_{min}}^{r(T)} dR (R-R_{min})^{\beta}\nonumber \\
       &= \frac{NC_0}{\beta +1}  (r(T) - R_{min})^{\beta + 1}\label{kuyf}  \\
&= \frac{NC_0 }{(\beta +1)} R_{min}^{\beta +1} \left( \frac{  T-T_{min}
                                            }{ T_m-T}                \right) ^{\beta +  1}, 
\end{align}
by use of \eq{distrjih} and \eq{jkhbrwa}.
Close to the absolute freezing point $T_{min}$ the above denominator is
close to $(T_m-T_{min})^{\beta +  1}$, so we shall use
\be
C(T) = B \left( \frac{T-T_{min}}{T_{min}} \right)^{\beta +1 }
\label{likjg}
\ee
where
\be
B= C_0 \phi  \left( \frac{ R^2_{min} }{ R_{max}-R_{min}}
 \frac{\rho_i \lambda T_{min}}{3 \sigma T_m}
  \right)^{\beta +1 }
\label{jhves}
  \ee
  by use of \eq{jkhbrwa} and the definition of
$N$. 

\subsection{Contribution of a pre-melted surface layers}             
Having neglected the thickness of the pre-melted films in the 
ice-filled pores by setting
$A_H=0$ we should compare the relative contributions to $C(T)$ from
these films and the liquid filled pores. Since there is no film in the
liquid filled pores, we need only take the $R>r(T) $ pores into
account. We take  the film contribution
to be given by the film thickness $d=R-r(T)$ as
\be  
\Delta C(T) =   C_0
\int_{r(T)}^{R_{max}}dR  g(R)
\frac{\Delta V(R)}{V(R)}
\ee  
where  the fraction 
$\Delta V/V \approx  3 d /{R}$ is the ratio of the film volume to the pore 
volume. Then
\be  
\Delta C(T) = 3d  N  C_0
\int_{r(T)}^{R_{max}}dR  
\frac{ (R-R_{min})^{\beta} }{R}
\ee   
where $R_{max}$ is the upper cut-off for $g(R)$.
When $\beta =1$ this integral is easily evaluated to give
\be  
\Delta C(T) = 3d(T)  N  C_0 \left( R_{max}-r(T) - R_{min} \ln \left(
\frac{R_{max}}{r(T)}
\right) \right) .
\ee  
Taking the $R_{max}$ term to dominate in this expression and  using
\eq{kuyf} the ratio becomes
\be
\frac{\Delta C}{C} \approx 6d(T) \frac{R_{max}}{(r(T) - R_{min})^2}
\label{uuh}
\ee
which may well be larger than one 
when $r(T) \gtrsim R_{min}$.

However, as we shall see below, it is the rates of change $dC/dT$ of
the volume fractions that are  important, not the absolute value of
$\Delta C/C$. 
The film thickness $d$ may be estimated from \eq{kiuiyg}  as the minimum 
of $\Delta \Omega$ when $\sigma =0$. This gives the standard 
expression  \cite{israelachvili11} 
\be 
d=\left( \frac{A_HT_m}{6\pi(T_m-T)\rho_i\lambda } \right)^{1/3}. 
\ee 
where we can use the relatively high value $A_H= 10^{-19}$ J. Together
with the constants given in table \ref{table} this gives $d\approx $1 nm,
while the other relevant length, which appears in \eq{kiuiyg},  is $  3\sigma/(\rho_i \lambda) \approx $ 0.3 nm.

Since $C \propto (T-T_{min})^2$ and, to leading order $\Delta C \propto d
\propto (T-T_m)^{-1/3}$, we have that $dC/dT = (\beta +1 ) C/ (T-T_{min}) $ while
$d\Delta C/dT =  \Delta C/(3 (T_m-T) )$, so that the ratio of the
changes in these two quantities due to a temperature change when $\beta
=1$ is
\be
\frac{\delta \Delta C}{\delta C} = 
\frac{T_m}{T-T_{min}} \frac{dr_0R_{max}}{R_{min}^3}.
\ee
close to the absolute freezing point $T_{min}$. Here we have used
\eq{jkhbrwa}
to substitute $(T_m-T_{min})/T_m = r_0/R_{min}$.
The condition that $\delta \Delta C /{\delta C} \ll 1$ may be taken as
a condition on the range of pore sizes $R_{max}/R_{min}$:
\be
\frac{R_{max}}{R_{min}} \ll \frac{R_{min}^2}{dr_0} \frac{T-T_{min}}{T_m} ,
\ee
or, equivalently,
\be
\frac{T-T_{min}}{T_m-T_{min}} \gg \frac{d}{R_{min}}\frac{R_{max}}{R_{min}} .
\ee
When $R_{min}= $ 30 nm for instance, and $(T-T_{min})/T_m =$ 1$/$273, we
get that $R_{max} /R_{min} \ll 10$.
It is quite natural that the condition for the domination of pore-
versus film fluid is a limited range of pore sizes, as a domination of the large pores,
that all carry a film contribution, would leave a smaller fraction of the
porosity to be represented by smaller pores. 

In other words, when  $R_{max} /R_{min} \ll 10$,
$\Delta C$ changes relatively slowly with $T \approx
T_{min}$,  but $C$ changes significantly. In this case we may neglect the variations
in the film contribution to the overall change in liquid volume
fraction.
For this reason we shall only use the  $C$ 
in the following, keeping in mind that it is the fraction of liquid
pore water, and not the total fraction of liquid water.

\subsection{Governing equation for the evolution of melted water concentration }

 In a 1d setting  the conservation of energy in a
slab of thickness $dx$ over a
time $dt$
may be written
\be
(j(x) - j(x+dx))\Delta Adt= \lambda \rho \Delta A dx dC 
+ c_b\Delta A dx dT 
\label{balance}
\ee
where $\Delta A$ is the cross-sectional area, $c_b$ the combined specific heat
capacity of the porous medium and the water,  and $T$ is the temperature. 
In \eq{balance} the left hand side is the net energy
transfer to the slab, the first term on the right is the energy consumed
by melting (latent heat), and the  last term, the energy absorbed
due to  the  heat capacities of the water, ice and the porous medium itself.
As $dx \rightarrow 0$ \eq{balance} becomes
\be
\frac{\partial j}{\partial x} + \lambda \rho \frac{\partial C}{\partial t} + 
c_b
\frac{\partial T}{\partial t} =0.
\label{balance2}
\ee
To describe the heat flow we apply  the Fourier law, which takes the form
\be
j = -\kappa
\frac{\partial T}{\partial x} 
,
\ee
where $\kappa$ is the bulk thermal conductivity of the porous medium, so
inserting this in
\eq{balance2}  gives
\be
\frac{\partial }{\partial x} 
 \left(    \kappa  \frac{\partial T}{\partial x}  \right)   =
\left( \rho \lambda \frac{\partial C}{\partial T} + c_b\right)
\frac{\partial T}{\partial t} 
\label{kjhgyt}
\ee
where we have used
${\partial C}/{\partial t} =({\partial C}/{\partial T})( {\partial
  T}/{\partial t} )$.
Generalizing  to 
arbitrary dimension ($\partial /\partial x \rightarrow \nabla$) and
replacing $T-T_{min} $ by $T_{min} (C/B)^{1/(\beta +1 )}$  using \eq{likjg}, 
yields the 
diffusion equation 
\be  
(1 + M) \frac{\partial   C}{\partial t}      =
D_0 
\nabla \cdot 
\left(    C^{1/(\beta +1 ) -1} \nabla C  \right)  .  
\label{kjhfgtwa}
\ee 
where 
\be 
D_0= \frac{\kappa_b T_{min} 
}{ \lambda \rho (\beta +1 ) B ^{1/(\beta +1 )}}
= \frac{c_b T_{min} 
}{ \lambda \rho (\beta +1 ) B ^{1/(\beta +1 )} } D_t ,
\ee 
and $D_t = \kappa_b/c_b \approx $ 1mm$^2/$s is the average thermal diffusivity 
of the porous  medium, and
\begin{align} 
M& = \frac{c_bT_{min} C^{1/(\beta +1 ) -1}}{(\beta +1 ) \rho_i \lambda
   B^{1/(\beta +1 )}} \nonumber \\
&= \frac{3c_b \sigma T_m }{(\beta +1 ) (\rho_i \lambda )^2 (C_0 \phi
  )^{1/(\beta +1 )}}
\frac{R_{max}-R_{min}}{R_{min}^2}
C^{-\gamma}.
\end{align}
 The last expression comes from replacing $B$ by the expression
in \eq{jhves}, and $ \gamma  = \beta /(\beta +1 ) $. 
Using the material constants in table \ref{table} gives
\be
M = l \frac{R_{max}-R_{min}}{R_{min}^2} C^{-\gamma}
\ee
where $l\approx$ 0.4 nm. 
 So $M\ll 1$ when the range of pore sizes is limited and  as long as
 $C$ does not become too small.

 So, we
shall proceed to analyze the case where the $M$-term may be dropped,
leaving the equation
\be 
\frac{\partial   C}{\partial t}      = 
D_0  
\nabla \cdot (   C ^{- \gamma }    \nabla  C  ) .
\label{jhytg}
\ee  
We note at this point that the condition for neglecting the energy
needed to change temperature, which is represented by the $M$-term,
coincides
with the condition to neglect the contribution  of pre-melted
films. Both conditions may be fulfilled by media with  a limited range
of pore sizes above a minimum size $R_{min} \gtrsim $10 nm. 

 The fact that we have   
neglected the energy contribution given by the heat capacities means
that we have assumed   
that all the energy is spent melting the ice in the pores. The mobile energy  
density $\propto C$, which means that \eq{jhytg} may be read as  
a statement of energy conservation.

We will consider the response  to a localized addition of energy
causing a local initial volume $V_i$ of melted water.
Solving \eq{jhytg} subject to the normalization condition 
\be 
V_i = \int dV C 
\label{norm}
\ee 
 in $d$ dimensions \cite{flekkoy21} for a point source initial
$  C (r,0) $ gives
\be
  C(r,t)
=\frac{ p(r/t^{\tau} ) }{f^{d}(t)}
\label{form}
\ee
where
\begin{equation}
    f(t)=\left(\frac{2-d\gamma}{(1-\gamma)V_i^{\gamma}}  D_0t\right)^{\frac{1}{2-d\gamma}}\;. 
\label{kjgyr}
  \end{equation}
and
\begin{equation}\label{eq11}
    p(y)=\left[\frac{\gamma}{2(1-\gamma)V_i^{\gamma}}y^2+k\right]^{-\frac{1}{\gamma}}
  \end{equation}
where $y=r/f(t)$ and $k$ is an integration constant given by the
normalization condition. It takes the value \cite{flekkoy21}
\begin{equation}
  k=\left[V_i^{1-d\gamma /2}\left(\frac{\gamma}{2\pi(1-\gamma
        )}\right)^{\frac{d}{2}}\frac{\Gamma\left(\frac{1}{\gamma}\right)}{\Gamma\left(\frac{1}{\gamma}-\frac{d}{2}\right)}\right]^{\frac{2\gamma}{d\gamma-2}}\; .
  \label{jkhyre} 
\end{equation}

The  functional form given in \eq{form} immediately yields the second moment for
the concentration  profile 
\be
r_{rms}^2  = \frac{\int dr r^{d-1} r^2    C(r,t)   }{\int dr r^{d-1}  
  C(r,t) }  \propto  f^{2}(t) \propto  t^{2\tau}
\ee
with \cite{pattle59,flekkoy21}
\be 
\tau = \frac{1}{2-d\gamma} , 
\ee
or, in terms of $\beta $
\be 
\tau = \frac{(\beta +1 ) }{d-(d-2)(\beta +1 )} , 
\label{uygkjh}
\ee
which  in 3d yields hyper-ballistic spreading ($\tau >1$)  if
$1/2 <\beta  < 2$ and super-diffusion ($\tau > 1/2$) if  $0<\beta  <2$.
A value of  $\beta =1$, which would correspond to a 
linear  initial growth of $g(R)$ gives $\beta  =1$ and $\tau =2$,
which should be compared to the ballistic value $\tau =1$, and
the normal diffusive value of $\tau =1/2$.
A heat pulse will thus spread at an accelerating rate, causing a
rapid long range front of melting.

The $d=1$ case  is relevant when a heat pulse  spreads
downwards in the ground. In this case $\beta =1$ gives $\tau=2/3$ and
$\beta =2$ gives $\tau=3/4$, both super-diffusive values. 

The super-diffusive spread of $C$ and thus $T$ with  time thus
follows from the fact that a heat pulse will lower  the local melting 
temperature, thus keeping the remaining ice from receiving more latent heat as the 
temperature is rising.  In contrast,  a melting front that propagates
through a medium with a single pore size will only spread diffusively,
propagating at a speed $\sim \sqrt{t}$.

\section{Potential application to  tundra-like surfaces}

On the tundra an increase in  heat penetration depth due to
super-diffusion, will enhance the water melting
caused by annual heating, thus increasing the melting depth. 
Freezing and melting on a tundra is believed to affect the subsurface over
depths of the order $\sim$4 m. Provided the relevant range of pore
sizes is present, we may speculate that when the ground is heated by
the sun,
melting fronts lasting   days or months, will propagate downwards giving rise to a
one-dimensional problem of the type we have discussed above.
How much deeper will a super-diffusive spreading of heat, or
$C$-fluctuations propagate than the  normal melting front, and will this have an effect
on the release of trapped methane?

It is instructive  first  to consider the case  of a medium with a single pore size
$R_{min}$, and look at the case where   a heat pulse propagates
from
the  surface. Then  $g(R)=\delta (R-R_{min} )$ 
and all the pores melt at the same temperature $T_{F0}$ 
given by setting $T\rightarrow T_{F0}$ in \eq{jkhbrwa}. Taking this to be the initial temperature in the
ground,  the  temperature will spread out downwards untill it
reaches the melting front
below which  $T = T_{F0}$. The volume fraction as a function of $T$ is
$C=C_0 \Theta (T-T_{F0} )$, where $\Theta$ is the Lorentz-Heaviside function, and
\be
\frac{dC}{dT} = C_0\delta (T-T_{F0} )
\ee
which is zero away from the melting front. In this case $dC/dT$ cannot
be assumed larger than $c_b$; rather, for $T\neq T_{F0}$ \eq{kjhgyt} reduces to the normal
diffusion equation
\be
  \frac{\partial T}{\partial t}  =
\frac{\kappa}{c_b}   \frac{\partial^2 T}{\partial x^2}  
\label{kjhgkjgyt}
\ee
which describes standard diffusive spreading of $T$.

At the point where all the energy supplied at the surface has been
consumed as latent heat at the melting front, the front propagation
stops. This will happen at a depth $z_f = Q/(\rho \lambda)\sim$ 1-4 m where $Q$
is the thermal energy per unit area initially supplied at the
surface and $\rho$ the total mass density. This layer is usually called the 'active layer'.

Now, returning to the  case we have considered, where $g(R)\propto  (R-R_{min} )^{\beta}$,
what happens in the one-dimensional case when a heat pulse propagates
from the surface and downwards?
This question may be answered by examining the analytic solutions
given in   \eq{form}. Choosing $\beta =1$ and 
setting $d=1$ in \eq{uygkjh}
gives the diffusion exponent $\tau = 2/3$. If  $\beta =2$, then  $\tau =
3/4$. These values are sufficiently close that it may
be hard to distinguish between them experimentally, and so the end
result will not depend strongly on the $\beta$-value. So, we shall
use $\beta =1$, which corresponds to a linear increase in $g(R)$ for
$R$ near $R_{min}$.

\begin{figure}[h!]
\begin{center}
\includegraphics[width=0.85\columnwidth]{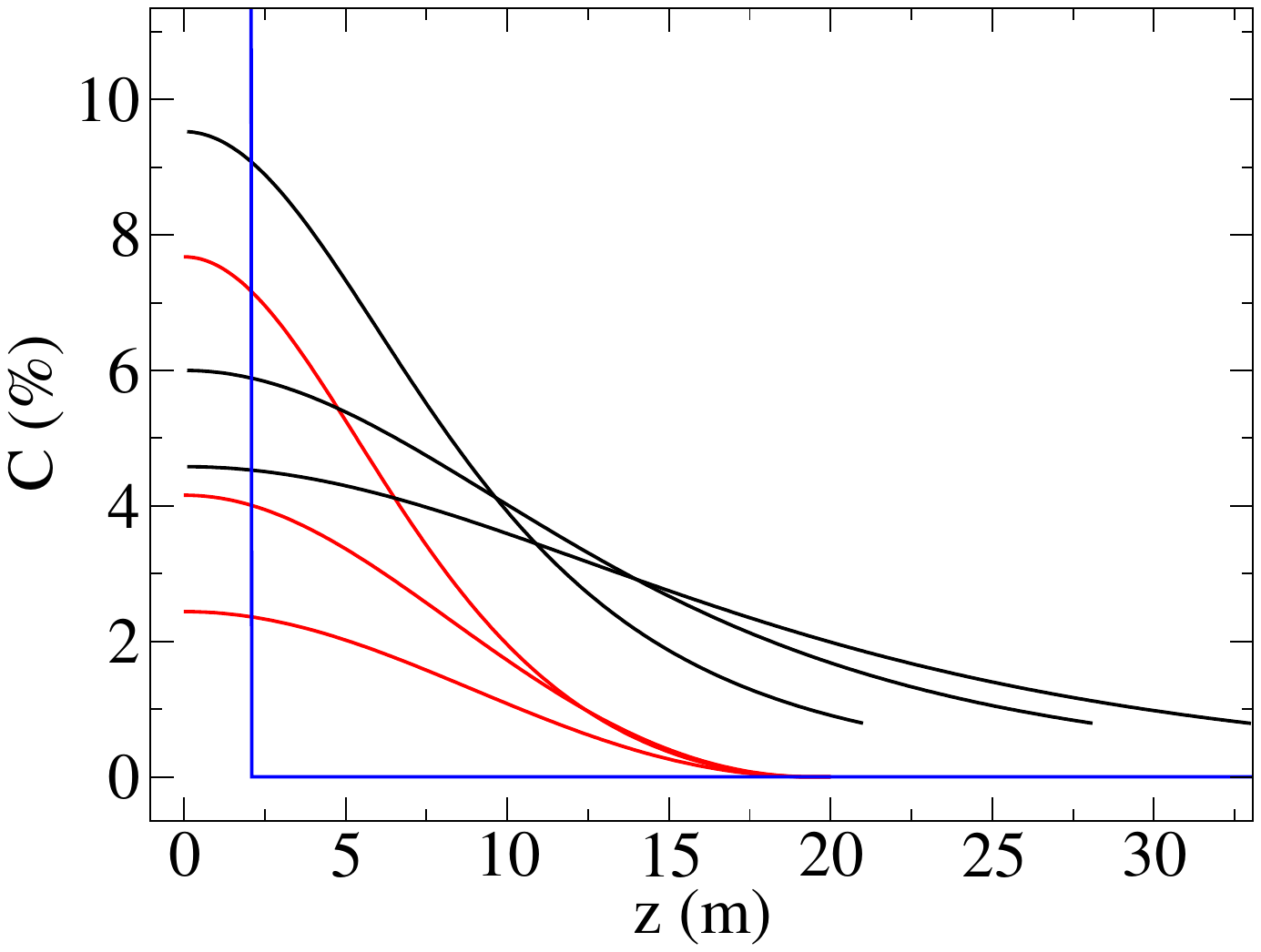}
\includegraphics[width=0.85\columnwidth]{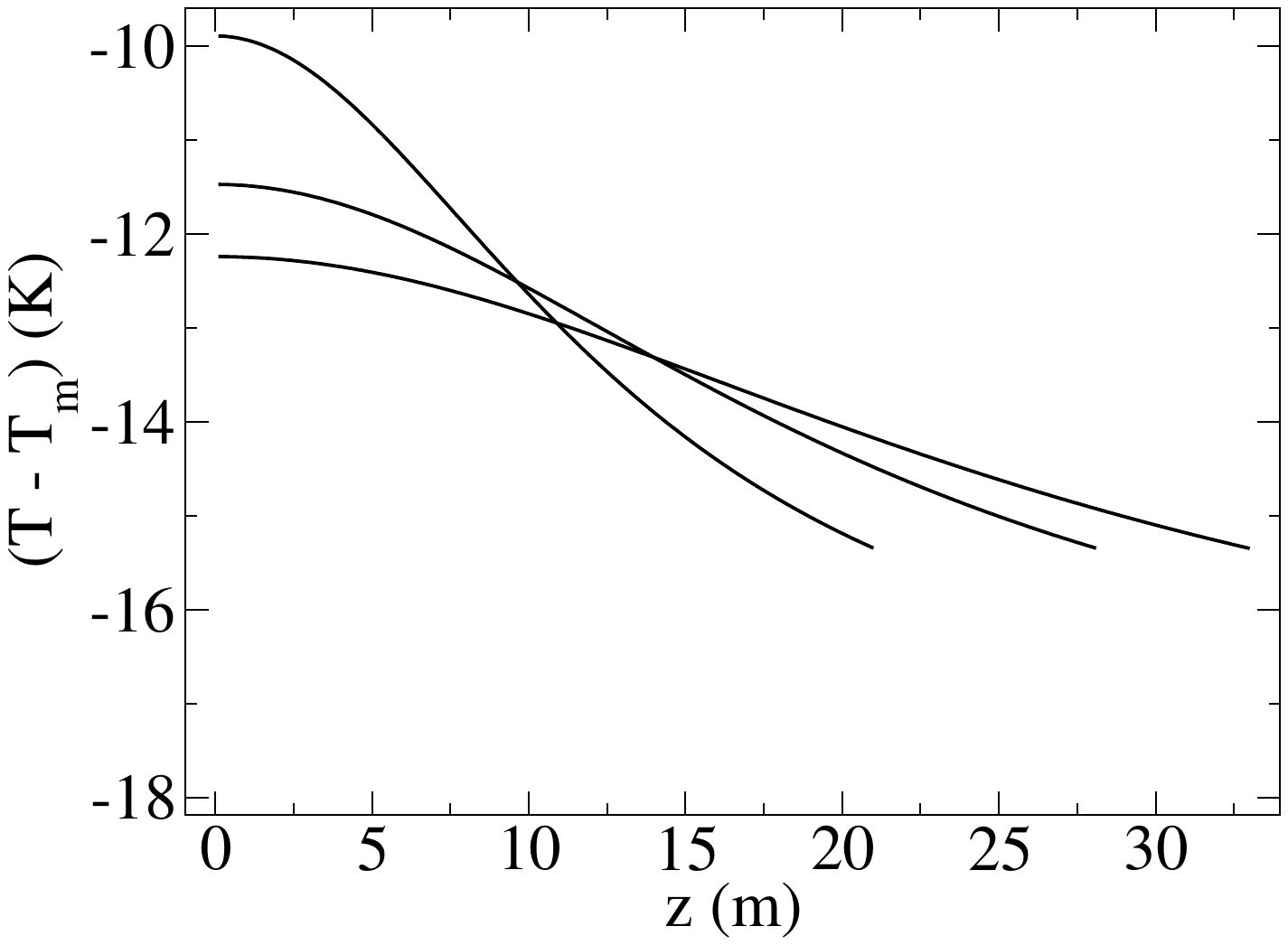}
\caption[]{Top: The melted water fraction as a function of depth at 
  different times $t=$4, 8, and 12 months. 
  The black curves show the analytic solutions of  \eq{kjhfgtwa} in the 
  domains where they are assumed to apply, that is, where $T<T_{max}$
  and $M<1$.  The red curves 
show the corresponding numerical solution of the full heat equation 
\eq{kjhfgtwa}. The blue 
curve shows the case where there is a constant pore size and no 
super-diffusive spreading.
Bottom: The corresponding temperature.
Here $\beta =1$, $R_{min}=R_{max}/2=$ 5 nm, and $c_b= 2 
$MJ$/$m$^3$
as is close to both the ice and typical clay/silt values. 
}
\label{jhgt}
\end{center}
\end{figure}  Using the  value of the  thermal diffusivity for ice $D_t \approx $ 1
mm$^2/$s and $R_{max}= 2R_{min}=$ 60 nm, which  yields $D_0\approx $ 0.015 mm$^2/$s, we can estimate the
typical penetration depths with the super-diffusive contribution of the latent heat:
In ref. \cite{flekkoy21} it is shown that
the second moment of $C$ 
is given by 
\begin{align} 
  r^2_{rms}
  & =\frac{d }{2 }  \pi^{\frac{d}{2}}k^{\frac{d}{2}+1-\frac{1}{\gamma}}
\frac{\Gamma\left(\frac{1}{\gamma}-\frac{d}{2}-1\right)}{\Gamma\left(\frac{1}{\gamma}\right)}
    \left(\frac{2(1- \gamma )}{\gamma}\right)^{\frac{d}{2}+1}
            \nonumber \\
&     V_i^{\frac{d\gamma}{2}+\gamma - 1}f^2(t) .
\label{jmhvyd}
\end{align}
The 1-d solution is given by setting $V_i=2 \phi C_0 l_d$,  where $l_d=$
is the initial thickness of the active (melted) layer,  and $\phi$ is the
porosity.
The factor 2 comes from the fact that our solution describes the symmetric situation where $C(z,t)$
spreads
out symmetrically in both directions from $z=0$, while we are
interested in the case where it only spreads downwards.
Setting $l_d=$ 2m, $C_0=1$, $\beta  =$1, $d=$1 the result becomes
\be  
r^2_{rms} = (2\pi )^{2/3} \left( \frac{3D_0 t}{(2 \phi C_0 l_d)^2}
\right)^{4/3} (2\phi C_0 l_d)^2.
\label{jkhuyf}
\ee  
Inserting the numbers $\phi=0.5$, $C_0=1$, $l_d=$ 2m
and $D_0=$ 0.01-0.1 mm$^2/$s the $r_{rms}$ may be written
\be 
r_{rms} \sim \left( \frac{t}{\mbox{year}} \right)^{2/3} \mbox{ 10 m}
\ee 
which is a typical factor 10 or so larger than $l_i$. This result
shows   that  super-diffusive spreading of heat may cause
temperature variations almost an order of magnitude deeper than the
variations caused by a normal diffusive melting front.

\begin{figure}[h!]
\begin{center}
\includegraphics[width=0.85\columnwidth]{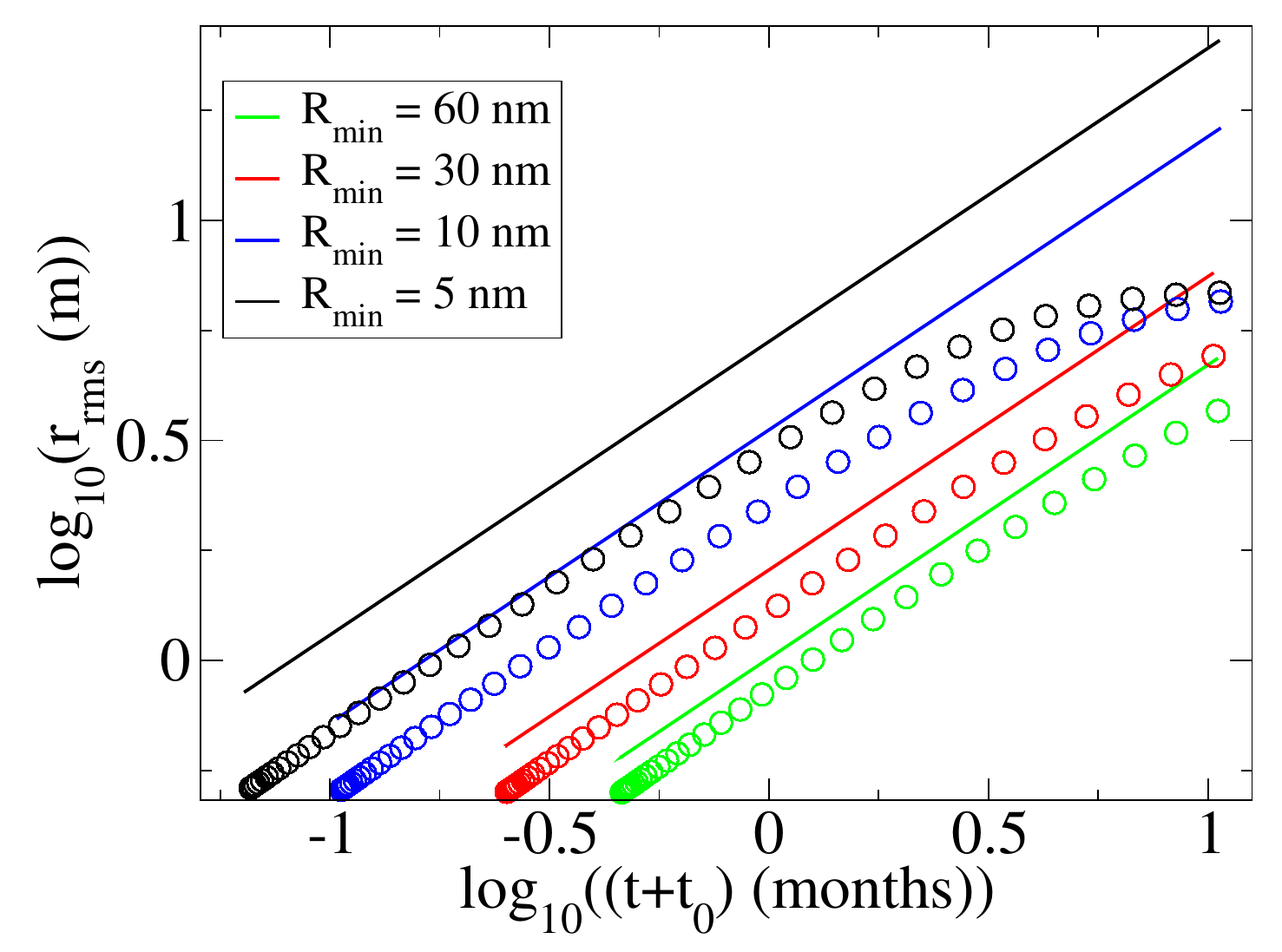}
\caption[]{The increase in  rms depth of the melted water fraction as function of
  time for different minimum pore sizes. The time increases  from $t=$ 0,
and $t_0>0$ corresponds to the fact that, for numerical reasons,  the initial $C(z,0)$-profile
is not a $\delta$-function, but has a finite width.
  The full lines, drawn in colors that correspond to the   ($\circ$)-points, show the theoretical value of
  \eq{jkhuyf} while the ($\circ$)-points  show the values measured from the
  numerical solution of \eq{kjhfgtwa}. Parameter values are as in figure \ref{jhgt}.}
\label{jhgiugt}
\end{center}
\end{figure}

The main approximation made in our theory is the neglect of the heat
capacities compared to the latent heat contributions.
We now solve the full heat balance equation \eq{kjhfgtwa}
numerically, including the finite value of the heat capacity:
\begin{figure}[h!]
\begin{center}
\includegraphics[width=0.85\columnwidth]{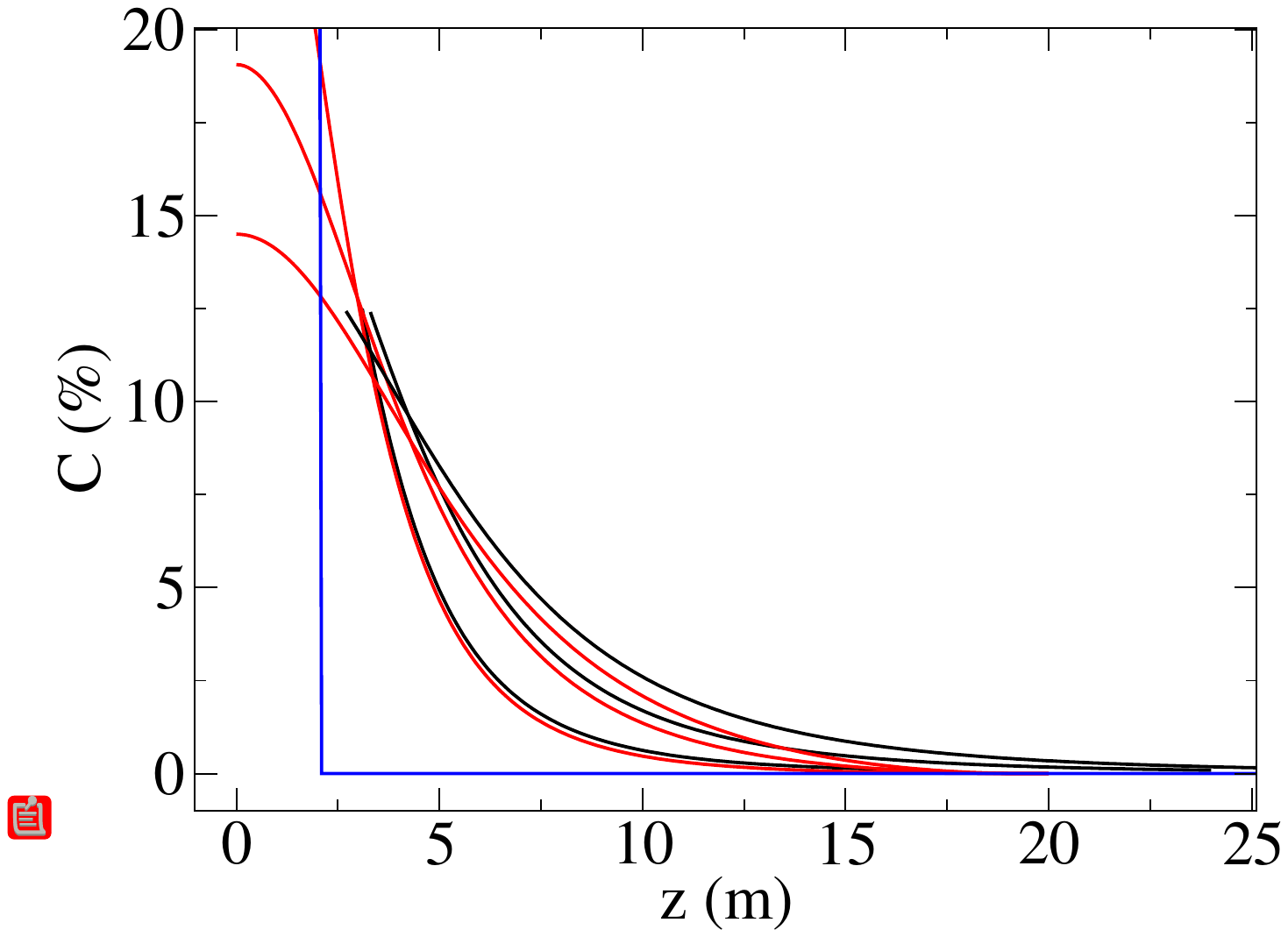}
\caption[]{The same results as in figure \ref{jhgt}, but with
 $R_{min}=R_{max}/2=$ 30 nm}
\label{jhgtreal}
\end{center}
\end{figure} 
Figure \ref{jhgt} shows the results of this. Note that the analytic solutions are only plotted for $C$
values where $M< 1$ and $T<T_{max}$, the freezing
temperature of pores of size $R_{max}$. It is seen that the full
solution of \eq{kjhfgtwa} gives a somewhat smaller $z$-value where $C$
approaches zero, but the scaling of $r_{rms}\propto t^{\tau}$ with time is
still seen to hold for the first months after the heat pulse, as may
be seen from figure \ref{jhgiugt}.
For larger $R_{min}$-values
 there is increasingly good agreement between the 
numerical solution and the analytic solution based on the $c_b=0$ approximation.
Even in the $R_{min}= $10-60 nm   cases the penetration of $C\gtrsim $
1\%  fluctuations extend deeper than 10 m as may be seen from figure \ref{jhgtreal}.

\section{Conclusions}

Starting from the thermodynamics of the Gibbs-Thomson effect
describing the melting of ice in pores and
a power law distribution of pore sizes we have shown that the
requirement of energy conservation   produces a non-linear equation that yields
super-diffusive spreading of the melted water fraction. 

The physical picture that emerges from this analysis is  that the
spreading of heat, or the melted water concentration,
is strongly enhanced by the fact that the heat will bypass any pore
that is either too big for melting to occur, or so small that the
melting has already happened. This is true in the range
of temperatures  where some pores contain water and some contain
ice. 
As a result, a subsurface porous medium containing ice, will experience
melting perturbations at depths that greatly exceed those that are
expected from a treatment that ignores the freezing point depression.

The super-diffusive spreading of temperature or melted water fraction,
may also be used as a method to   measure pore size distributions:
 The estimate given in \eq{uuh} shows that  close to $T_{min} $, the sensitivity to
 temperature variations is mainly in the pore
 liquid fraction $C$, and not the liquid fraction contained in the
 surface melted films. In the cases where the pore size distribution
 is in  fact given by a power law distribution, the
 measurement of a spreading temperature profile may thus provide a value
 for the diffusion exponent $\tau $ and thus  for the pore size
 distribution exponent $\beta$.
 Due to the higher sensitivity to the bulk pore water,
 this method may be superior to conventional NMR measurements which cannot
 distinguish  between the liquid water that resides in the pores and that which is
 contained in the films. Compared to mercury injection measurements,
 which need high pressures to probe the smallest pores, the temperature
 technique is less likely to alter the medium through crushing of the
 smallest pores. It does however, rely on the basic assumption of a
 power law pore size distribution.

Finally, we note that methane hydrates, which may exist in the
subsurface where glaciers have recently withdrawn, have similar
values of density and latent heat as water ice \cite{chuvilin18}. This
may give rise to super-díffusive behavior, even when the active
substance is not water, but methane in combination with water.
Measurements showing freezing point depression of methane and CO$_2$
hydrates in natural sediments \cite{park18}  supports  this
assertion. 

\begin{acknowledgements}
We thank Daan Frenkel for valuable suggestions on the significance of
the nucleation effects.
  This work was partly supported by the 
Research Council of Norway through its Centers of Excellence funding 
scheme, project number 262644. AH acknowledges funding from the
European Research Council (Grant Agreement 101141323 AGIPORE).
\end{acknowledgements}

\end{document}